\documentclass[webconf, onecolumn]{svjour}
\usepackage{graphics}
\usepackage[varg]{txfonts} 
\usepackage[latin1]{inputenc}
\session-title{Adaptive Optics for Extremely Large Telescopes}
\def\arcsec{\nobreak{$''$}}

\begin{document}
\title{Science and Adaptive Optics Requirements of MICADO, the E-ELT
  adaptive optics imaging camera}
\titlerunning{Science and AO requirements of MICADO}
\author{Richard~Davies\inst{1}\fnmsep\thanks{\email{davies@mpe.mpg.de}} 
\and
N.~Ageorges\inst{1}
\and
L.~Barl\inst{1}
\and
L. Bedin\inst{10}
\and
R.~Bender\inst{1,3}
\and
P.~Bernardi\inst{8}
\and
F.~Chapron\inst{8}
\and
Y.~Clenet\inst{8}
\and
A.~Deep\inst{4}
\and
E.~Deul\inst{4}
\and
M.~Drost\inst{6}
\and
F.~Eisenhauer\inst{1}
\and
R.~Falomo\inst{7}
\and
G.~Fiorentino\inst{5}
\and
N.~M.~F\"orster~Schreiber\inst{1}
\and
E.~Gendron\inst{8}
\and
R.~Genzel\inst{1}
\and
D.~Gratadour\inst{8}
\and
L.~Greggio\inst{7}
\and
F.~Grupp\inst{3}
\and
E.~Held\inst{7}
\and
T.~Herbst\inst{2}
\and
H.-J.~Hess\inst{3}
\and
Z.~Hubert\inst{8}
\and
K.~Jahnke\inst{2}
\and
K.~Kuijken\inst{4}
\and
D.~Lutz\inst{1}
\and
D.~Magrin\inst{7}
\and
B.~Muschielok\inst{3}
\and
R.~Navarro\inst{6}
\and
E.~Noyola\inst{1}
\and
T.~Paumard\inst{8}
\and
G.~Piotto\inst{7}
\and
R.~Ragazzoni\inst{7}
\and
A.~Renzini\inst{7}
\and
G.~Rousset\inst{8}
\and
H.-W.~Rix\inst{2}
\and
R.~Saglia\inst{1}
\and
L.~Tacconi\inst{1}
\and
M.~Thiel\inst{1}
\and
E.~Tolstoy\inst{5}
\and
S.~Trippe\inst{1,9}
\and
N.~Tromp\inst{6}
\and
E.~A.~Valentijn\inst{5}
\and
G.~Verdoes~Kleijn\inst{5}
\and
M.~Wegner\inst{3}
}
\authorrunning{R. Davies and the MICADO Phase A Team}
\institute{Max Planck Institute for extraterrestrial Physics, Germany
\and 
Max Planck Institute for Astronomy, Germany
\and
Munich University Observatory, Germany
\and
University of Leiden, Netherlands
\and
University of Groningen, Netherlands
\and
NOVA Optical/IR Instrumentation Group, Netherlands
\and
Astronomical Observatory of Padova, INAF, Italy
\and
Laboratory of Space Studies and Instrumentation in Astrophysics, 
Observatory of Paris, France
\and
Institute for Millimetre Radio Astronomy, France
\and
Space Telescope Science Institute, USA
}

\abstract{
MICADO is the adaptive optics imaging camera being studied for the E-ELT.  
Its design has been
optimised for use with MCAO, but will have its own SCAO module for the
initial operational phase; and in principle could also
be used with GLAO or LTAO. 
In this contribution, we outline a few of the science drivers for MICADO
and show how these have shaped its design. The science drivers have
led to a number of requirements on the AO system related to
astrometry, photometry, and PSF uniformity. We discuss why these
requirements have arisen and what might be done about them.
} 
\maketitle

\section{MICADO Overview}
\label{sec:micado}

MICADO is the {\em M}ulti-AO {\em I}maging {\em Ca}mera for {\em D}eep 
{\em O}bservations, which is being designed to work with adaptive
optics on the 42-m European Extremely Large Telescope.
During the Phase A study, the instrument has been optimised for the
multi-conjugate adaptive optics module MAORY \cite{dio09};
but it is also able to work with other adaptive optics systems.
In particular, it includes a separate module to provide a single
conjugate adaptive 
optics capability with natural guide stars during the early
operational phase. 
Combined with the simplicity and robustness of the camera design,
which have been borne in mind throughout its development, this has led
the consortium to view MICADO as an ideal E-ELT first light
instrument.

\begin{figure}
\centering{\resizebox{0.8\columnwidth}{!}{\includegraphics{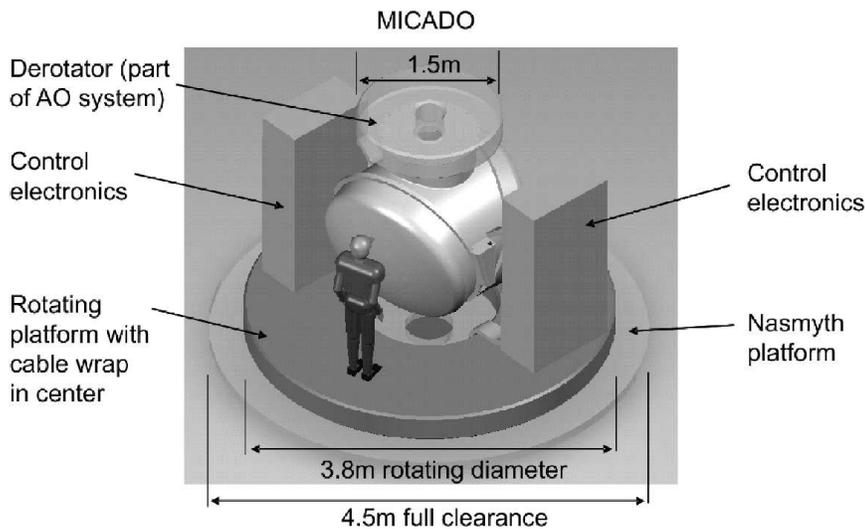} }}
\caption{View of MICADO as it will hang underneath the AO systems. The
  cryostat is less than 2m across, and all key components on the
  inside can be accessed through two large doors without dismounting
  the instrument. The electronics will be in two cabinets on a
  co-rotating platform, which also houses the cable-wrap and itself
  sits directly on the Nasmyth platform.}
\label{fig:micado}
\end{figure}

The instrument is very compact and is supported
underneath the AO systems so that it rotates in a
gravity invariant orientation, as shown in Fig.~\ref{fig:micado}.
It is able to image, through a large number of selected
wide and narrow band near infrared filters, a wide 53\arcsec\ 
  field of view at the diffraction limit of the E-ELT. 
In addition to a high throughput camera with a fixed 3\,mas pixel
scale, MICADO will have an auxiliary arm to provide an increased
degree of flexibility. 
In the current design, this arm provides 
(i) a finer 1.5\,mas pixel scale over a smaller field, and
(ii) a 4\,mas pixel scale for a simple, medium resolution longslit
spectroscopic capability.
However, in principle the auxiliary arm also opens the door to many
other options, including simple polarimetry, a `dual imager' based on a
Fabry-Perot etalon to image separate emission line and continuum
wavelengths simultaneously, or a high time resolution detector.

\section{MICADO Key Capabilities}

Early in the project, the consortium highlighted several key
capabilities that exemplify the unique features of the E-ELT, and at
which MICADO on the E-ELT will excel in comparison to other
facilities. 
These are at the root of the science cases \cite{ren09} and have
driven the design of the camera.
\\

{\bf Sensitivity and Resolution.}
Its high quality optics and 3\,mas pixels mean MICADO is optimised
for imaging at the diffraction limit, and will fully sample the
6--10\,mas FWHM in the J--K bands.
With a throughput exceeding 60\% its sensitivity at 1--2\,$\mu$m will,
for the AO performance predicted by MAORY, be comparable to, and may
even surpass, JWST.
But MICADO's superior resolution means that it will be able to
probe the detailed structure of objects that are unresolved by JWST.
In addition, its field of view of nearly 1\,arcmin yields a
significant multiplex advantage compared to other ground-based cameras
such as IRIS on the TMT.
Together, these characteristics make MICADO a powerful tool
for many science cases, from studies of faint high redshift galaxies
to performing photometry in crowded fields.
This latter topic is discussed in more detail in Sec.~\ref{sec:phot}.
\\

{\bf Precision Astrometry.}
The primary imaging field of MICADO employs a catoptric design using
only fixed mirrors.
Together with the gravity invariant rotation and the baseline to use
Hawaii-4RG detectors (developed to meet the stringent
requirements of space astrometry missions), this makes MICADO an ideal
instrument for astrometry.
A robust pipeline, based on software already available in the
AstroWISE system, will bring precision astrometry into the mainstream.
An analysis of the statistical and systematic effects \cite{tri09}
shows that proper motions of 40\,$\mu$as/yr in a single epoch of
observations should be achievable.
At this level, many novel science cases become feasible, and a few of
these are discussed in Sec.~\ref{sec:astrom}.
\\

{\bf High Throughput Spectroscopy.}
The obvious complement to plain imaging is spectroscopy, and this
capability is mandatory if MICADO is to be considered as a first light
instrument.
We have implemented a simple slit spectrometer with a high throughput
that is ideal for obtaining spectra of compact objects.
The resolution of $R\sim3000$ is sufficient to probe between the near
infrared OH lines.
This simple addition will enhance many science cases, for example: 
deriving stellar types and 3D orbits in the Galactic Center; 
using velocities of stars in nearby galaxies to probe central black
hole masses and extended mass distributions; 
measuring absorption lines in galaxies at $z=2$--3 and emission lines
in galaxies at $z=4$--6 to derive their ages, metallicities, and star
forming histories;
and obtaining spectra of the first supernovae at $z=1$--6.
\\

{\bf Simple, Robust, Available Early.}
Imaging is fundamental to astronomy, and MICADO is designed to excel
at this technique.
By doing so it will both exploit and promote the most unique features
of the E-ELT.
The instrument's optical and mechanical simplicity lead directly to
the stability needed for astrometry and photometry;
and in addition mean that the instrument can be available for first
light.
In order to capitalize on this potential, MICADO has been designed to
work not only with MCAO, but also has its own robust SCAO system
that will provide wavefront sensing with NGS to control
the telescope's deformable and tip-tilt mirrors.

\section{Astrometry}
\label{sec:astrom}

With an astrometrical precision of 40\,$\mu$as and proper motions of
10\,$\mu$as/yr detectable after only a few years, one finds that
`everything moves' and many remarkable science cases become feasible.
MICADO will be able to measure stellar orbits not only in the Galactic
Center, but also around other supermassive black holes in nearby
galaxies.
The proper motions of globular clusters can be used to derive their
distance via parallax displacement, separate their stellar populations
from field stars, and, by looking for kinematic families, probe the
formation and evolution of the Galaxy.
By analysing the internal kinematics of dwarf spheroidals to derive
their anisotropy and mass density profiles, one can test 
hierarchical models of structure formation.
Perhaps one of the most fascinating topics that can be
resolved is that of intermediate mass black holes.

\subsection{Intermediate Mass Black Holes}

Black holes with masses of order $10^3$--$10^4$\,M$_\odot$ are
expected theoretically from simulations of massive dense star
clusters, such as the Arches cluster \cite{por06} and IRS\,13
\cite{mai04} near the centre of the Galaxy.
In the most massive globular cluster Omega-Cen, the mass has been
measured as $4\times10^4$\,M$_\odot$ \cite{noy08}.
However, recently doubts have been cast on this result from
astrometric data, which have been use to derive proper motion dispersions
along both the radial and tangential directions.
They have revealed a small but significant degree of
anisotropy \cite{and09,mar09}.
For a luminosity profile with a shallow cusp, the authors found 
M$_{\rm BH}\sim2\times10^4$\,M$_\odot$; but remarkably their model with a
core profile was consistent with no black hole.
Thus these authors could only put an upper limit on the black hole mass, and
the question of whether Omega-Cen contains a black hole is once again open.

This result was based on 4 years of HST imaging which yielded
an astrometric precision of about 100\,$\mu$as.
MICADO will be able to achieve better precision in a single epoch, and
hence opens the way to probing the anisotropy and black hole masses
in this and many other globular clusters.

\subsection{Statistical and Systematic Effects}

There are many effects that have to be taken into consideration in the
instrument design, and carefully calibrated, in order to reach the
astrometric precision required \cite{tri09}.
Of these, 3 are relevant to adaptive optics correction.
Differential tilt jitter is the small residual motion of objects in
the focal plane, and although an order of magnitude smaller for MCAO
than for SCAO, is still significant. 
Because it is a statistical effect, it has to be integrated out,
although more advanced methods of combining frames can lead to better
results \cite{cam09}.
Both instrumental and atmospheric effects can influence the NGS used
for tip-tilt sensing.
For example, if rotation of the instrument is inexact, the NGS will
appear to move in different directions.
Alternatively both achromatic refraction (which arises because
opposite sides of the field of view are at different airmass) and
chromatic refraction (which depends on the colour of the NGS) will
also cause the NGS to drift slowly. 
As the AO system corrects for these shifts, it will warp the field of
view.
If the warping is, as expected, only a low order effect, it can easily
be calibrated out in individual frames.
On the other hand, it may be possible to avoid it completely by
updating the centering position of the NGS on their detectors to follow
their time-averaged barycentres.
Spatial variations and asymmetries in the PSF could also lead to
astrometric inaccuracies.
This is harder to quantify without detailed and time comsuming PSF
simulations; but work on preliminary analytical PSFs by the MAORY consortium
\cite{dio09} suggests that it should have little impact.
The importance and impact of PSF knowledge -- or lack of it -- is
discussed in the next section.

\section{PSF and Photometry}
\label{sec:phot}

Most science cases require at least some knowledge of the PSF.
In the following subsections we look at 2 highly contrasting science
cases that both require a good knowledge of the PSF, but apply it in
different ways.

\subsection{QSO Host Galaxies}

\begin{figure}
\centering{\resizebox{0.75\columnwidth}{!}{\includegraphics{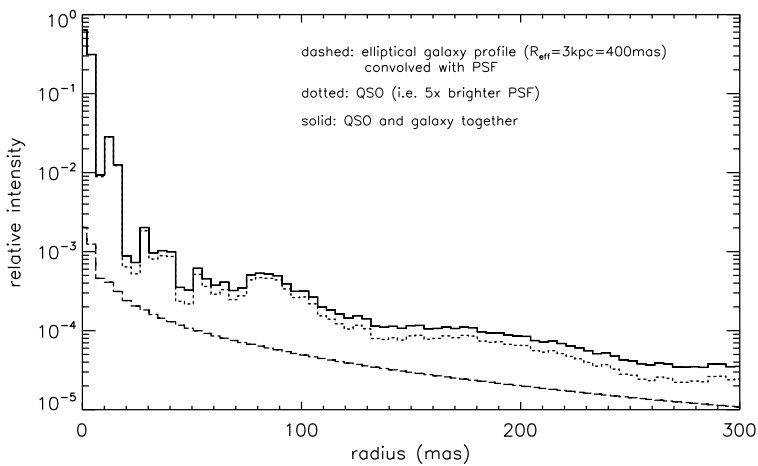} }}
\caption{Illustration of a QSO and its host galaxy at high redshift
  ($z>2$), showing the
  similarity between the radial profiles of the two components; and
  hence that in order to subtract the QSO contribution to reveal the
  underlying host galaxy, one needs to know the PSF well (a simple
  Strehl ratio is insufficient: a detailed profile is required).}
\label{fig:qso}
\end{figure}

Some of the key issues in galaxy evolution concern the relation
between the mass of a supermassive black hole in the centre of a galaxy
and that of the stellar spheroid surrounding it, the mechanisms that
regulate their respective growth, and the rapid truncation of star
formation that leads to the observed colour bimodality.
In our current understanding, it is AGN feedback that is responsible
for all these. 
As such, QSOs and their host galaxies represent a crucial piece of the
puzzle, and studying these to high redshift is an important goal.
However, measuring the properties of the host galaxy requires one to
subtract the central QSO, which is often several to tens of times
brighter -- a task that is increasingly difficult at higher redshifts,
and requires a good estimate of the PSF \cite{jah04}.
This is illustrated in Fig.~\ref{fig:qso}, which shows the radial
profile one might expect for an elliptical host galaxy, the radial
profile of the central QSO (i.e. a PSF), and the small difference
between them.
Approximately, for a QSO that is 5 times brighter than the host
galaxy, an error of only 2\% in the estimation of the Strehl ratio
leads to an uncertainty of 10\% in the host galaxy magnitude.
And if the error on the Strehl is as much as 20\%, then one might not
even detect the host galaxy at all.

From the astronomer's perspective, for science cases such as this,
reconstruction of the on-axis PSF from WFS data would the ideal solution.
Yet good PSF reconstruction, specifically when one or more LGS are
used in the wavefront sensing, remains a challenge.
In this respect the large field of view of MICADO is an asset even
though the science target itself is compact, because
it becomes highly likely that suitable empirical PSFs can be
found within the field even at high galactic latitudes.
If the PSFs do not vary much across the field, they can be combined
and used directly.
Alternatively if one can parametrize several PSFs in different
locations it may be possible to interpolate the PSF at the position of
the QSO, or indeed at any other point in the field.
A positive step in addressing this by developing a simple model
for the PSF has been taken by the MAORY consortium \cite{dio09}.

\subsection{Resolved Stellar Populations}

An alternative way to study galaxy evolution is through the relic
populations in nearby galaxies \cite{tol09}.
Accurate photometry of spatially resolved stellar populations enables
one to create colour-magnitude diagrams in order to derive a galaxy's
star formation history. 
The ultimate goal of such work is to apply these methods to galaxies
in the Virgo 
cluster which, at a distance of $\sim$17\,Mpc, is the nearest large
cluster, with over 2000 member galaxies of all morphological types.
Directly counting the number of stars on different branches enables
one to derive a coarse but robust star formation history for a galaxy
that, by probing the oldest stars with ages $>10^{10}$\,Gyr, can be
traced back to the early universe \cite{gre02}.
In the Leo Group (at a distance of $\sim$11\,Mpc), MICADO can
comfortably reach the Horizontal Branch.
In closer galaxies, such as Cen\,A at $\sim$3\,Mpc, it will be possible
to measure the entire Red Giant Branch (RGB) down to the level of the
main sequence turnoff, and hence to address the problem of the
age-metallicity degeneracy.

\begin{figure}
\centering{\resizebox{0.9\columnwidth}{!}{\includegraphics{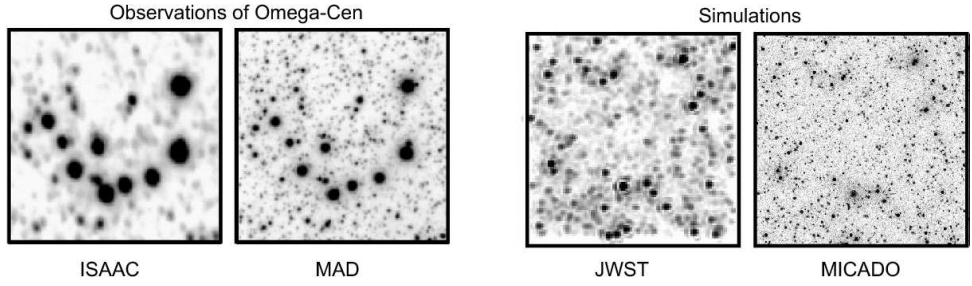} }}
\caption{Examples showing how, by reducing crowding, enhanced
  resolution can increase the effective sensitivity. 
Left: observations of Omega-Cen by ISAAC and MAD at the VLT. 
Right: simulations of a crowded field as seen by JWST and MICADO. In
  crowded fields, MICADO's higher resolution will enable it to reach
  several magnitudes deeper than JWST.}
\label{fig:crowding}
\end{figure}

Although sensitivity is important in such work, the crucial aspect is
resolution since this allows one to probe ever more crowded fields, as
illustrated by Fig.~\ref{fig:crowding}.
Observations with ISAAC and MAD at the VLT of the sme stellar field
have shown that the improvement in resolution from 0.6\arcsec\ to
0.1\arcsec\ leads to 3\,mag greater depth \cite{mar07,bon09}:
by reducing the crowding, better resolution yields an effective
gain in sensitivity.
The factor 6 in resolution between ISAAC and MAD is similar to the
difference between JWST and MICADO, and so we also expect to reach several
magnitudes deeper in crowded fields.
It means that, in the Virgo cluster, rather than probing the fringes
of galaxies at 3--4\,R$_{\rm eff}$, we will be able to work near
the centres of the galaxies at 1--2\,R$_{\rm eff}$.
This is a fundamental issue, since it enables one to study the regions
of galaxies where a large fraction of the stellar mass is located, and
hence gain significant information about the global star formation
history. 
In addition it will be possible to address the problem of stellar
population gradients in giant galaxies using photometry of individual
stars, rather than integrated light profiles.

Despite the science field being full of stars, one still needs to be
careful about extracting the PSF from the highly crowded data.
A number of widely used programs already exist to do this, such as
DAOPHOT and StarFinder, and these appear to work well also on simulated
MICADO frames.
But one aspect that needs to be assessed further is the impact on
photometric accuracy of spatial variations across the large field.
Simulations show that if these are not taken into account, the photometric
scatter increases markedly.
However, one possible solution is to make use of the fact that in
MICADO the field is imaged onto an array of 16 detectors, where each
detector subtends an angle of only 12.3\arcsec.
And on any single detector, PSF variations will be negligible.
Thus, until spatially variant PSFs can be dealt with in photometry
packages, an interim way to cope with them is simply to treat each
detector independently.

\section{Conclusion}
\label{sec:conc}

MICADO is the adaptive optics imaging camera being studied at Phase~A
for the E-ELT.
It has been designed to be simple, robust, and could be ready for the
E-ELT first light.
The primary imaging field has a high throughput, reflective, gravity
invariant design using only fixed mirrors, making it ideally suited to
photometric and astrometric science applications.
But in order to achieve its goals, there are a two important adaptive
optics issues that need to be fully understood and resolved:
(i) the impact of instrumental and atmospheric effects on the NGS, and
what the knock-on effects are on field warping;
(ii) how best to acquire a detailed knowledge of the PSF and its
spatial variations.

%

%

%

\end{document}